\documentclass[twocolumn,preprintnumbers]{revtex4}
\usepackage{amsmath,amssymb}
\usepackage{graphicx}
\usepackage{dcolumn}
\usepackage{bm}

\begin{document}

\title{The combination of de Broglie's Harmony of the Phases and Mie's
theory of gravity results in a Principle of Equivalence for
Quantum Gravity.}

\author{E. Paul J. de Haas}
\homepage{http://www.physis.nl} \affiliation{epjhaas@tiscali.nl}

\date{{\it Les Annales de la Fondation Louis de
Broglie} {\bf 29} 707-726, (2004)}

\begin{abstract} Under a Lorentz-transformation, Mie's 1912
gravitational mass behaves identical as de Broglie's 1923
clock-like frequency. The same goes for Mie's inertial mass and de
Broglie's wave-like frequency. This allows the interpretation of
de Broglie's "Harmony of the Phases" as a "Principle of
Equivalence" for Quantum Gravity. Thus, the particle-wave duality
can be given a realist interpretation. The "Mie-de Broglie"
interpretation suggests a correction of Hamilton's variational
principle in the quantum domain. The equivalence of the masses can
be seen as the classical "limit" of the quantum equivalence of the
phases.
\end{abstract}

\maketitle

\section{Mie's foundations for a theory of matter.}

In 1912-1913 Gustav Mie published his "Grundlagen einer Theorie
der Materie" in a series of three papers in "Annalen der
Physik"\cite{Mie1},\cite{Mie2},\cite{Mie3}. His concern was the
failure of classical physics, mechanics and electrodynamics, in
the sub-atomic domain. Mie tried to find a fundamental connection
between the existence of quantized matter and the fact of gravity.
He used the Hamiltonian method in his attempt to elucidate the
existence of the electron, the quantum of action and the facts of
gravity. He reduced the problem of finding a theory of matter to
the problem of finding a universal function $\mathcal{H}$. This
$\mathcal{H}$ should be a function of the electromagnetic field,
of the electric potentials and of the gravitational field. And
this function, the universal Hamiltonian, should be invariant
under a Lorentz transformation. The latter ensured the invariance
of the formula Mie derived from this Hamiltonian. Mie assumed the
existence of such a universal Hamiltonian and then searched its
true expression.

Mie interpreted gravitation as a cohesive effect inherent in
energy as such. In the theory of relativity the inertial
energy-density $\mathcal{E}_i$ entered as the last part of the
stress-energy tensor, so he had to enter the complete tensor in
the equations. But then other four-dimensional quantities should
be incorporated as well and this led to such a level of
mathematical complexity that Mie concluded it, in 1912, to be
impossible to find a theory of gravity in this way. If, however,
he used the Lorentz-invariant Hamiltonian energy-density
$\mathcal{H}$ instead of $\mathcal{E}_i$ as having an inherent
cohesive effect, then Mie considered it not too difficult to
achieve the goal of finding a theory of gravity. Because
$\mathcal{H}$ was a Lorentz-scalar and because we have $d
 V = \frac{1}{\gamma}d V_0$, for a rest-system volume $V_0$ and
a volume in a moving system $V$ with
\begin{equation}\label{gamma}
\gamma = \frac{1}{\sqrt{(1-\frac{v^2}{c^2})}},
\end{equation}
Mie could combine this and conclude
\begin{equation}\label{H}
    \int_V \mathcal{H}d V = \frac{1}{\gamma}E_0
\end{equation}
and
\begin{equation}\label{Ez}
    \int_V \mathcal{E}_id V = \gamma E_0
\end{equation}
for a moving particle (\cite{Mie3}, p. 26). Mie interpreted
$\mathcal{E}_i$ as defining the inertial mass and $\mathcal{H}$ as
defining the gravitational mass (\cite{Mie3}, p. 40):
\begin{equation}\label{Ei}
   m_i c^2 = \int_V \mathcal{E}_id V = \gamma E_0
\end{equation}
\begin{equation}\label{mg}
   m_g c^2 = \int_V \mathcal{H}d V = \frac{1}{\gamma}E_0
\end{equation}
with
\begin{equation}\label{mi}
   m_i = \gamma m_0
\end{equation}
\begin{equation}\label{m}
   m_g = \frac{1}{\gamma}m_0.
\end{equation}

Mie concluded that movements of matter influenced gravitational
and inertial mass. Especially hidden movements of the elementary
particles inside matter, heat, caused the inertial mass to
increase and the gravitational mass to decrease (\cite{Mie3}, p.
49). In his theory of gravity the Newtonian principle of
equivalence, NEP or $m_{0i}=m_{0g}$, nowadays called the weak
equivalence principle or WEP, only applied to a particle in its
rest-system and became invalid in a moving frame. So the NEP was
not Lorentz-invariant and could therefore not function as an axiom
in his attempt to relativize gravity, dependent as it was on the
relative motion of the observer. Mie didn't come up with an
alternative principle of equivalence and he wasn't able to develop
his theory any further. This motivated Einstein to ignore Mie's
theory \cite{Norton}.

\section{Louis de Broglie's Harmony of the Phases.}

Ten years after the publication of Mie's papers, modern
post-orbital or post-"Bohr-Sommerfeld" quantum mechanics began
with de Broglie's hypothesis of the existence of matter waves
connected to particles with inertial mass. De Broglie started with
the assumption that every quantum of energy $E$ should be
connected to a frequency $\nu$ according to
\begin{equation}\label{hf}
E=h\nu
\end{equation}
with $h$ as Planck's constant \cite{deBroglie1},\cite{deBroglie2}.
Because he assumed every quantum of energy to have an inertial
mass $m_o$ and an inertial energy $E_0 = m_0 c^2$ in its
rest-system, he postulated
\begin{equation}\label{h0f}
h\nu_0=m_0 c^2.
\end{equation}
De Broglie didn't restrict himself to one particular particle but
considered a material moving object in general \cite{deBroglie1}.
This object could be a photon (an atom of light), an electron, an
atom or any other quantum of inertial energy. If this particle
moved, the inertial energy and the associated frequency increased
as
\begin{equation}\label{Eb}
h\nu_i=E_i=\gamma E_0=\gamma m_0 c^2=\gamma h \nu_0
\end{equation}
so
\begin{equation}\label{Ee}
\nu_i=\gamma \nu_0.
\end{equation}
But the same particle should, according to de Broglie, be
associable to an inner frequency which, for a moving particle,
transformed time-like in the same manner as the atomic clocks with
period $\tau_{atom}$ and frequency $\nu_{atom}$ do in Einstein's
Special Theory of Relativity. We quote Arthur Miller from his 1981
study on Einstein's Special Theory of Relativity (\cite{Miller},
p. 211). In this quote, the rest frame is named k and the moving
frame K. \begin{quote} In 1907 Einstein [.] defined a clock as any
periodic process -for example, an atomic oscillator emitting a
frequency $\nu_0$ as measured in k. [.]..an observer in K measures
the frequency:\end{quote}
\begin{equation}\label{te} \nu_{atom}=
\frac{1}{\gamma} \nu_0.
\end{equation}
\begin{quote} [.]the clock at k's origin registers a time observed
from K of:\end{quote}
\begin{equation}\label{period}
\tau_{atom}= \gamma \tau_0.
\end{equation}
Einstein attributed a clock-like frequency to every atom. De
Broglie generalized Einstein's view by postulating that every
isolated particle with a rest-energy possessed a clock-like
frequency. Thus, de Broglie gave every particle two, and not just
one, frequencies, their inertial-energy frequency $\nu_i$ and
their inner-clock frequency $\nu_{c}$. The inner-clock frequency,
of atoms and photons, was postulated by Einstein, the
inertial-energy frequency was postulated by de Broglie. These
frequencies were identical in a rest-system but fundamentally
diverged in a moving frame according to
\begin{equation}\label{innu}
\nu_i=\gamma \nu_0
\end{equation}
\begin{equation}\label{conu}
\nu_{c}= \frac{1}{\gamma} \nu_0.
\end{equation}
This constituted an apparent contradiction for de Broglie, but he
could solve it by a theorem which he called "Harmony of the
Phases". He assumed the inertial energy of the moving particle to
behave as a wave-like phenomenon and postulated the phase of this
wave-like phenomenon to be at all times equal to the phase of the
inner clock-like phenomenon. Both inner-clock- and wave-phenomenon
were associated to one and the same particle, for example an
electron, a photon or an atom. The inertial wave associated with a
moving particle not only had a frequency $\nu_i$ but also a
wave-length $\lambda_i$ analogous to the fact that any inertial
energy $E_i$ of a moving particle had a momentum $p_i$ associated
to it. De Broglie used the four-vector notation to generalize the
connection of a particles inertia to the associated
wave-phenomenon (\cite{deBroglie2}, Chap. II.5). This allowed him
to incorporate the momentum $\mathbf{p}_i$ and the wave-number
$\mathbf{k}_i$:
\begin{equation}\label{hrpc}
P_{\mu}= (\mathbf{p}_i, \frac{i}{c}E_i)=
h(\frac{1}{2\pi}\mathbf{k}_i,\frac{i}{c}\nu_i)= h O_{\mu}.
\end{equation}
The phase $\varphi_i$ of the wave-like inertial energy-momentum
four-vector $P_{\mu}$ became
\begin{equation}
\varphi_i=2\pi(\nu_it-\frac{1}{2\pi}\mathbf{k}_i\cdot \mathbf{r})=
- 2\pi O_{\mu}R^{\mu}
\end{equation}
or, in energy-momentum expression
\begin{equation}
\varphi_i = - \frac{2\pi}{h}(E_it - \mathbf{p}_i \cdot
\mathbf{r})= - \frac{1}{\hbar} P_{\mu}R^{\mu}
\end{equation}
which gave
\begin{equation}\label{hrp}
\hbar \varphi_i = - P_{\mu}R^{\mu}.
\end{equation}
De Broglie could show that his postulates ensured the law of the
Harmony of the Phases, the inertial wave-like phase equaling the
inner clock-like phase of the particle
\begin{equation}\label{harmonyphases}
\varphi_i=\varphi_c.
\end{equation}
The proof of the principle of equivalence of the phases is based
upon the Lorentz-transformation properties of four-vectors,
especially the invariance of the inner product,
\begin{equation}\label{hrpe}
\varphi_i = - 2\pi O_{\mu}R^{\mu} = -2\pi O_{0}R^{0}= 2\pi \nu_0
t_0,
\end{equation}
and the transformation-properties of the inner clock-like
frequency $\nu_c$ and the time-coordinate $t$
\begin{equation}\label{hrpf}
\varphi_c = 2\pi\nu_c t = \frac{1}{\gamma}2\pi\nu_0 \gamma t_0 =
2\pi\nu_0 t_0.
\end{equation}

The relativistic expressions for the inertial phase of a moving
particle allowed de Broglie to postulate a wave-length $\lambda_i$
associated to the magnitude of the electrons inertial momentum
$\mathbf{p}_i$
\begin{equation}\label{wavelength}
\mid \mathbf{p}_i \mid  = \frac{h}{\lambda_i}.
\end{equation}
This inertial momentum could be interpreted as generated by an
inertial energy-flow $E_i\mathbf{v}_{group}$ with
\begin{equation}\label{Eflow}
\mathbf{p}_i=\frac{E_i}{c^2}\mathbf{v}_{group}.
\end{equation}
The Harmony of the Phases resulted in a super-luminous
wave-velocity $v_{wave}$ connected to the particle-velocity
$v_{particle}$ as
\begin{equation}\label{w}
v_{wave} = \frac{c^2}{v_{particle}},
\end{equation}
but this was not in contradiction with the postulates of
Einstein's Special Theory of Relativity because the wave couldn't
carry energy and the group-velocity of the wave, $v_{group}$,
equalled the velocity of the associated particle, $v_{particle}$.
So the group velocity was connected to the moving inertial energy.

At first, these postulates were regarded as too fantastic to be
true. But Einstein recognized it as important and reported it to
the German physicist community. This allowed Schr\"{o}dinger to
use the ideas of de Broglie and in January 1926 he published his
famous wave-equation based upon the postulates of de Broglie. The
next year Davisson and Germer, working in the Bell lab in New
York, obtained the first electron diffraction pattern by
bombarding a crystal of nickel with a mono-velocity electron beam.
The theoretical incorporation and experimental confirmation of the
wave-aspect of particles with inertial mass prompted the general
acceptance of this part of de Broglie's ideas. The interpretation
of de Broglie's postulates soon became a central problem of the
fast developing quantum theory. However, in the battle into which
the interpretation problem of quantum physics transformed, the
idea of an inner, clock-like frequency associable to an electron
as a particle disappeared from the scene. All attention got
focussed on the nature of the matter waves connected to the
inertial energy (\cite{Sievers}, p. 27). Then in the final
interpretation of the Copenhagen School, the moving electron
completely evaporated in the wave and the inertial wave
transformed into an abstract probability wave disconnected from
physical reality
 \cite{deBroglie3}.

\section{De Broglie's inner frequency and Mie's gravitational energy.}

In the orthodox, Copenhagen School interpretation of quantum
physics, the moving electron was only represented by its
probabilistic wave aspect. When the moving "electron"-wave was
stopped in an experiment by placing a photographic plate in its
path, the wave mysteriously "collapsed" or vanished and the
particle miraculously reappeared as a dark spot on the
photographic plate. The wave was no longer seen as an inertial
wave, as in de Broglie's original papers, but as defining a
probability density connected to the prediction of experimental
outcomes, such as the chance of finding a dark spot on a
particular area of the photographic plate. The particle-wave
duality for moving particles, considered as a fundamental aspect
of physical reality by de Broglie, was "resolved" by cancelling
the particle-aspect and by interpreting the wave as an abstract
mathematical entity used to predict the outcome of experimental
setups. All references to an underlying physical reality in which
particles and waves had a real existence, were carefully expelled
from the theory \cite{deBroglie3}. This interpretation reflected
the philosophical spirit dominating the scientific circles of the
time, logical positivism in the line of Mach and the {\it Wiener
Kreis}. Einstein and de Broglie opposed this particular
philosophy, they inclined to common sense realism searching an
explanation for the mysteries of nature in terms of models
representing physical reality. Einstein and de Broglie wanted to
retain the physical reality of both waves and particles, and they
declared that
\begin{quote}
the formal concepts of the "orthodox" theory, while no doubt
giving precise statistical representations, did not present a
complete picture of physical reality \cite{deBroglie4}.\end{quote}
They did not succeed in formulating a viable alternative for the
interpretation of the Copenhagen School. However, if we connect
Mie's theory to de Broglie's, an interesting realist
interpretation arises, an interpretation that we can connect to a
more modern approach based on relativistic tensor-dynamics.

Our association of Mie with de Broglie starts with the observation
that de Broglie didn't connect a physical energy to the inner,
clock-like frequency of the electron. He proved the Harmony of the
Phases by using the wavelength and frequencies, not by means of
the momentum and energies. We ask ourselves what kind of energy we
should identify with the space left empty by de Broglie in the
following sentence: the wave frequency belongs to the inertial
energy of the particle as the inner clock-frequency belongs to the
....... energy of the particle. The answer that will link his
approach to Mie's is: the gravitational energy. If every quantum
of energy has to be connected to a frequency, as de Broglie
successfully postulated, then gravitational energy $E_g$ has a
gravitational frequency $\nu_g$ with
\begin{equation}\label{gravfreq}
E_g = h \nu_g.
\end{equation}
If we connect the relativistic Hamiltonian part of Mie's theory of
gravity to the basic postulates of the relativistic
quantum-frequency theory of de Broglie, we get
\begin{equation}\label{HMBa}
  \int_V \mathcal{H}d V = \frac{1}{\gamma}E_0 = E_{gravity}= h
    \nu_{gravity}.
\end{equation}
This allows us to identify $\nu_{gravity}$ with $\nu_{clock}$
because
\begin{equation}
h\nu_c=\frac{1}{\gamma}h\nu_0 = \frac{1}{\gamma}E_0 =h\nu_g
\end{equation}
so
\begin{equation}
\nu_{gravity} = \nu_{clock}.
\end{equation}
The clock-like frequency belongs to an inner aspect of the
particle, so we can associate gravity to the particle and inertia
to the wave. Our interpretation results in a real particle-wave
duality, because
\begin{equation}\label{HMBab}
  \int_V \mathcal{H}d V = \frac{1}{\gamma}E_0 = E_{gravity}= h
    \nu_{gravity}= h\nu_{particle}
\end{equation}
and
\begin{equation}\label{EMB}
   \int_V \mathcal{E}_id V = \gamma E_0 = E_{inertial}= h
    \nu_{inertial}= h\nu_{wave}.
\end{equation}

\begin{figure}
  \includegraphics[width=7cm]{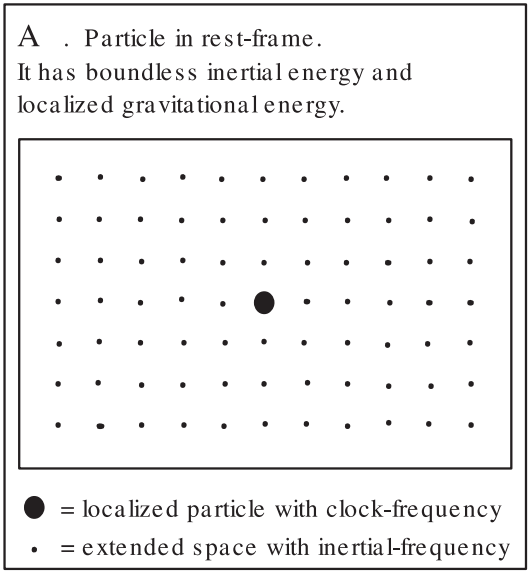} \\
  \caption{De Broglie's extended inertial frequency-field.}\label{duala}
\end{figure}

\begin{figure}
  \includegraphics[width=7cm]{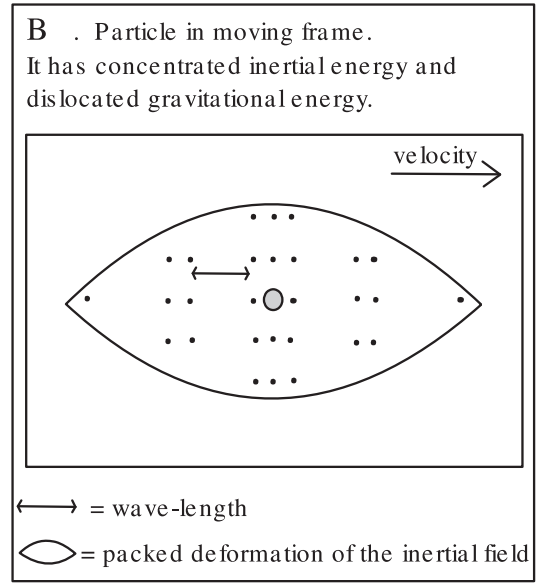} \\
  \caption{De Broglie's contracted inertial frequency-field.}\label{dualb}
\end{figure}

We will try to clarify this interpretation with the help of two
strongly simplified figures A and B. The particle with its inner
clock-like frequency is connected to the gravitational energy and
gravitational mass. The inner clock-like frequency $\nu_c$ could
also be called the particle-frequency $\nu_p$ or the frequency of
the gravitational energy $\nu_g$. In its rest-frame the
gravitational mass is concentrated at the place of the particle,
but in the moving frame the gravitational mass $m_g$ decreases and
becomes dislocated in the wave-packed. In its rest-frame, the
space around the particle is interpreted as an inertial field with
an inertial energy and a connected inertial frequency
$\nu_i=\nu_0$. For the particle at rest, this inertial field is
extended over the entire space so its density becomes infinitely
small and can't be measured. There is no wave-length because the
frequency and the connected energy are homogeneously spread out
over all of space. But when the particle moves, the inertial field
becomes inhomogeneous, acquires a wavelength $\lambda_i$ and the
inertial mass $m_i$ becomes concentrated in a small area
"surrounding" the dislocated particle. The inertial frequency
$\nu_i$ can now be called the wave-like frequency $\nu_w$ and
makes a four-vector with the wave-number
$\frac{1}{2\pi}\mathbf{k}_i$ or $\frac{1}{2\pi}\mathbf{k}_w$. This
inertial wave in its four-vector representation was written by de
Broglie as $O_{\mu}$ but is usually given in terms of the "angular
velocity" $\omega_i$ and the wave-number $k_i$ with
$K_{\mu}=(\mathbf{k}_i , \frac{i}{c}\omega_i)$ and $K_{\mu}= 2\pi
O_{\mu}$. De Broglie's association of a frequency-field to the
inertial energy Lorentz-transforms into the four-vector relation
for the inertial energy as a longitudinal wave $P_{\mu}= \hbar
K_{\mu}$. This inertial wave can be connected to the gravitational
energy, the latter being a fundamental property of the associated
particle.

Our interpretation puts the problem of the particle-wave duality
in a new perspective. The problem of localizing the particle in
the wave becomes a problem in the context of a theory of quantum
gravity. In this interpretation the problem of the
particle-localization is centered on the problem of the
localization of the gravitational energy within the
inertial-energy wave. The connection of the particle to the wave
and the connection of the gravitational mass to the inertial mass
are coinciding problems. The key to the enigma of the
particle-wave duality lies in de Broglie's "Harmony of the
Phases". In a Mie-de Broglie theory of Quantum Gravity, the
Newtonian principle of equivalence of the masses
\begin{equation}\label{eqmass}
m_g = m_i
\end{equation}
is valid only in a rest-frame of a particle. It becomes invalid in
a moving frame or for a moving observer. In a quantum context it
has to be replaced by the principle of equivalence of the phases
\begin{equation}\label{eqfase}
\varphi_g=\varphi_i.
\end{equation}
In this interpretation, the connection of the gravitational energy
to the inertial energy according to de Broglie's "Harmony of the
Phases" becomes the dynamical heart of quantum theory. The pilot
wave interpretation of de Broglie can be re-evaluated in this new
perspective. A particle moving, on a macroscopic scale,
"uniformly" through space deforms the metric on a quantum local
scale with its gravitational and inertial masses. In the process,
the inertial energy flow $E_i\mathbf{v}_{group}$ becomes
concentrated in a wave packed and the gravitational energy flow
$E_g\mathbf{v}_{particle}$ becomes dislocated in this wave packed.
From a macroscopical point of view, the two energies still
coincide. On a quantum scale, the attempt of de Broglie to
formulate a pilot-wave theory, in which a real particle is guided
by a real wave on its real path in space (\cite{deBroglie5}, p.
185-186), can, in the perspective of our present interpretation,
only be accomplished in a fully developed theory of Quantum
Gravity. Our pilot wave becomes the compressed inertial field or
the quantum local deformed metric and the particle trajectory must
be identified with the delocalized world tube of the gravitational
energy within this deformation. This strongly and not
coincidentally matches Vigier's description of his approach to
unify general relativity and quantum mechanics (\cite{Vigier}, p.
200).

\section{The Harmony of the Phases, the Hamiltonian and relativistic dynamics.}

Using the interpretation of the previous section it is quite easy
to prove de Broglie's postulate of the Harmony of the Phases. If
$\varphi_g=\varphi_i$ in the rest-frame and if they should remain
equal when the frame is set in motion we must have
\begin{equation}
d\varphi_g = d\varphi_i.
\end{equation}
For an infinitely small boost of the reference frame we can assume
the energy-momentum to be unchanged and set
\begin{equation}
\hbar d\varphi_g = E_g d t = \frac{1}{\gamma}E_0 \gamma d t_0 =
E_0 d t_0
\end{equation}
and
\begin{equation}
\hbar d\varphi_i = - P_{\mu}d R^{\mu}= E_i d t - \mathbf{p}_i d
\mathbf{r}= E_0 d t_0
\end{equation}
So our interpretation makes the postulate of the Harmony of the
Phases rather trivial.

We can modernize our interpretation further by showing that the
Hamiltonian of Mie's theory is not that outdated as it seems. We
can connect Mie's source of gravitational energy to the modern
one, the trace of the inertial stress-energy tensor
\cite{Norton},\cite{Rindler}. A free moving particle with inertial
mass $m_i$, momentum-density $g_i$ and four-momentum density
$G_{\nu}$ has an inertial stress-energy tensor
\begin{equation}\label{vg}
T_{\mu\nu}=V_{\mu}G_{\nu}
\end{equation}
with, in a (+,+,+,-) metric,
\begin{equation}\label{tracevg}
Trace(T_{\mu\nu})= V_{\mu}G^{\mu}= \mathbf{v}\cdot \mathbf{g}_i-
\mathcal{E}_i.
\end{equation}
For Mie's universal Hamiltonian we have
\begin{eqnarray}\label{hminvg}
    \int_V (\mathcal{H}-\mathcal{E}_i)d V =
    (\frac{1}{\gamma}-\gamma)E_0 \\
    = (1-\frac{v^2}{c^2}-1)\gamma E_0
    = -\frac{v^2}{c^2}E_i = -\mathbf{v}\cdot \mathbf{p}_i
\end{eqnarray}
and so (\cite{Mie3}, p. 52)
\begin{eqnarray}\label{minvg}
\int_V \mathcal{H}d V =\int_V \mathcal{E}_id V + \int_V
(\mathcal{H}-\mathcal{E}_i)d
V  \\
= E_i - \mathbf{v}\cdot \mathbf{p}_i = - V_{\mu}P^{\mu},
\end{eqnarray}
which gives
\begin{equation}\label{Htrace}
\mathcal{H} = - V_{\mu}G^{\mu}= - Trace(T_{\mu\nu})=
\mathcal{E}_g.
\end{equation}
[In Mie's treatment, the identification of $\mathcal{H}$ with $-
V_{\mu}G^{\mu}$ depends on a specific state of internal
oscillating motion of the system in consideration (see
\cite{Mie3}, p. 52 for further details).]

Ultimately we can relate the Harmony of the Phases to a very
general tensor equation. If we define the action tensor as
$S_{\mu\nu}= P_{\mu}R_{\nu}$ and use a four-volume $d\tau = d x d
y d z d ict $, then we have the invariant relation
\begin{equation}\label{tensorST}
T_{\mu\nu}= \frac{\partial ic S_{\mu\nu}}{\partial\tau}.
\end{equation}
We can write this as a differential equation
\begin{equation}\label{diffTS}
T_{\mu\nu}d \tau = d ic S_{\mu\nu}
\end{equation}
and with the approximation
\begin{equation}\label{approxPR}
d P_{\mu}R_{\nu}\approx P_{\mu}d R_{\nu}
\end{equation}
we get an equation that could well be the Harmony of the Phases in
differential form and in its tensor generalization,
\begin{equation}\label{tensorphase}
T_{\mu\nu}d \tau = ic P_{\mu}d R_{\nu}.
\end{equation}
If we concentrate on the trace of both sides we have
\begin{equation}\label{tracediff}
V_{\mu}G^{\mu}d \tau = ic P_{\mu}d R^{\mu}.
\end{equation}
With $\hbar d \varphi_i = - P_{\mu}d R^{\mu}$ and $d \varphi_i= d
\varphi_g$ we must have
\begin{equation}\label{tracephase}
 - V_{\mu}G^{\mu}d \tau = ic\hbar d\varphi_g,
\end{equation}
an equation which we interpret as the Harmony of the Phases in its
differential tensor trace expression. We can write it as an
integral equation
\begin{equation}\label{intTS}
\hbar\varphi_g = - \frac{1}{ic} \int_{\tau}V_{\mu}G^{\mu} d \tau =
- \int_{R}P_{\mu}d R^{\mu}= \hbar\varphi_i.
\end{equation}
The left side parts of this expression incorporate the particle
aspect and the right side parts the wave aspect. The right side
wave aspect of equation (\ref{intTS}) can be connected to
Sommerfeld's quantum rule for the phase integral or scalar action
$S$, (see \cite{Sommerfeld}, p. 102):
\begin{equation}\label{Jintegral}
- S = - \int_{R}P_{\mu}d R^{\mu} = n_k \hbar,
\end{equation}
with $n_k$ as the quantum number connected to the k-th degree of
freedom. According to Sommerfeld, the occurrence in nature of a
minimum variation  in the action ($\delta S$), connected to the
jump of a quantum system from one phase-state to the next, was the
fundamental reason for the appearance of Planck's constant
$\hbar$. According to Sommerfeld (\cite{Sommerfeld}, p. 97), the
minimum variation of the phase integral, or scalar action $S$, was
\begin{equation}\label{Sintegral}
- \delta S = - \delta \int_{R}P_{\mu}d R^{\mu} = \hbar.
\end{equation}

The left side particle aspect of equation (\ref{intTS}) can be
rewritten as
\begin{equation}\int_{\tau} \frac{\mathcal{H}}{ic} d
\tau = \hbar \varphi_g.
\end{equation}
This can be connected to Mie's application of Hamilton's principle
and to de Broglie's identification of Fermat's principle with the
principle of least action of Maupertius. Mie formulated his
relativistic version of Hamilton's variational principle ($\delta
S=0$) as (\cite{Mie1}, p. 527)
\begin{equation}\label{varprincMie}
\delta \int_{\tau} \frac{\mathcal{H}}{ic} d \tau = 0
\end{equation}
This relativistic version of Hamilton's variational principle was
called {\it Mie's axiom of the world function ($\mathcal{H}$)} by
Hilbert and was transformed by the same Hilbert into a general
covariant variational principle. Hilbert's version was assimilated
by the theorists of general relativity \cite{Vizgin}. But in our
Mie-de Broglie theory of Quantum Gravity, we have to correct Mie's
use of Hamilton's variational principle. A minimum variation of
the action integral can't be zero any more, because we have
\begin{equation}
 \delta\int_{\tau} \frac{\mathcal{H}}{ic} d \tau = \delta \hbar  \varphi_g
\end{equation}
and the quantum minimum of variation is one unit of action
$\hbar$, when $\delta \varphi_g =1$. This means that the Mie-de
Broglie version of Hamilton's variational principle in the quantum
domain should be
\begin{equation}\label{Qvarprinc}
 \delta\int_{\tau} \frac{\mathcal{H}}{ic} d \tau = \hbar.
\end{equation}
In the classical limit, on a scale where $\hbar\approx 0$, we get
the usual non-quantum version of Hamilton's principle. This
explains the failure of Mie to repair the breakdown of classical
physics, where the minimum variation of the action-integral is
assumed to be zero, in the quantum domain, where the minimum
variation equals Planck's constant $\hbar$. So the Harmony of the
Phases, or the principle of equivalence of the phases, as
expressed in equation (\ref{intTS}), leads to a very fundamental
correction of Mie's use of Hamilton's variational principle in the
quantum domain ($\delta S=\hbar$ instead of $\delta S=0$).

\section{On the relativity of the principle of equivalence.}

We thus "derived" the postulate of the Harmony of the Phases from
the tensor-equation (\ref{tensorST})
\begin{equation}\label{STtensor}
T_{\mu\nu}= \frac{\partial ic S_{\mu\nu}}{\partial\tau}.
\end{equation}
We will now investigate under which circumstances we may apply the
equivalence of the masses and/or the equivalence of the phases.
Let's assume an observer in a reference frame K to be capable of
measuring the values for $R_{\mu}$ and $P_{\nu}$ within a certain
degree of accuracy. Then he is able to calculate the associated
action tensor $S_{\mu\nu}$ and the stress-energy tensor
$T_{\mu\nu}$ unambiguously. This allows him to define the inertial
mass and the gravitational mass as in Mie's theory as
\begin{equation}
   m_i c^2 = \int_V \mathcal{E}_id V = - \int_V T_{44} d V
\end{equation}
\begin{equation}
   m_g c^2 = \int_V \mathcal{H}d V =
   -\int_V T^{trace}_{\mu\nu} d V.
\end{equation}
This observer is able to check the conditions under which the
inertial mass equals the gravitational mass (the weak principle of
equivalence). He will set
\begin{equation}
  \int_V T^{trace}_{\mu\nu} d V = \int_V  T_{44}d V,
\end{equation}
which gives
\begin{equation}\label{LaueStatic}
  \int_V (T^{trace}_{\mu\nu}- T_{44}) d V = 0
\end{equation}
and
\begin{equation}
  T^{trace}_{\mu\nu} - T_{44}= \mathbf{v}\cdot \mathbf{g} - \mathcal{E}_i +
  \mathcal{E}_i
  = \mathbf{v}\cdot \mathbf{g}= 0,
\end{equation}
with finally
\begin{equation}\label{vgPressure}
   \mathbf{v}\cdot \mathbf{g} = 0
\end{equation}
as the condition for which $m_i=m_g$. Because $\mathbf{v}\cdot
\mathbf{g}$ equals the pressure $p$, this implies a pressureless
situation as a necessary condition for the equivalence of
gravitational and inertial masses. The observer in K will conclude
that the equivalence of the masses is not a Lorentz-invariant
condition and cannot be transformed into a fundamental axiom or
law of nature. The same observer can define the inertial phase and
gravitational phase as
\begin{equation}
\hbar\varphi_i = - \int_{R}P_{\mu}d R^{\mu}
\end{equation}
and
\begin{equation}
\hbar\varphi_g =  - \frac{1}{ic} \int_{\tau} T^{trace}_{\mu\nu} d
\tau.
\end{equation}
The observer is able to check the conditions under which the
inertial phase equals the gravitational phase. He will find
equation (\ref{STtensor}) as the necessary condition. He will
conclude that the equivalence of the phases is a Lorentz-invariant
condition and that this equivalence can be seen as a fundamental
law of nature. If this observer 1 wants to communicate his
findings to a second observer in reference frame K', he must first
instruct observer 2 to measure $R_{\mu}'$ and $P_{\nu}'$ within a
certain degree of accuracy in K'. Then observer 2 must follow an
identical procedure as observer 1, but with the values of his own
K' in order to check the conditions for $m_i'=m_g'$ and
$\varphi_i'=\varphi_g'$. He will find the conditions to be
\begin{equation}
   \mathbf{v}'\cdot \mathbf{g}' = 0
\end{equation}
for the first and
\begin{equation}
T_{\mu\nu}'= \frac{\partial ic S_{\mu\nu}'}{\partial\tau'}
\end{equation}
for the second. Both observers will conclude that there are only
particular reference frames in which the inertial mass equals the
gravitational mass but that the inertial phase equals the
gravitational phase in every reference frame. The particular set
are the reference frames in which the pressure vanishes. The whole
procedure is in accordance with the principles of relativity and
not a single absolute value or reference frame is introduced.

We can connect this issue to a discussion between Schr\"{o}dinger
and Einstein regarding the stress-energy tensor for the universe
as a whole. In 1918 Schr\"{o}dinger argued that a stress-energy
tensor with a trace $T= -p-p-p+(\rho-p)= 0$ was possible as a
solution for the Einstein Equations, with negative pressure $-p$
and density $\rho$ \cite{Schrodinger}. Einstein answered to have
considered such a possibility but that he rejected it because this
negative pressure couldn't vanish in free space, which would
implicate a negative mass distribution throughout interstellar
space, a concept that meant the negation of free space
\cite{Einstein}. In his General Theory of Relativity, Einstein
mainly considered situations in which the total pressure $p$
vanished and the only non-zero component of the stress-energy
tensor was $T_{44}$. This matched Laue's condition for completely
static systems, by which Laue meant all systems for which equation
(\ref{LaueStatic}) holds (\cite{Norton}, p. 58). So Laue's and
Einstein's completely static systems are exactly those for which
Mie's definitions give $m_i=m_g$. We quote John Norton from his
1992 study of the emergence of Einstein's gravitational theory
(\cite{Norton}, p. 58):
\begin{quote} Notice that Einstein can only say he does justice to
the equality of inertial and gravitational mass "up to a certain
degree", since this result is known to hold only for completely
static systems and then only in their rest frame.
\end{quote}
We conclude that Mie's definitions of $m_i$ and $m_g$ are in
accordance with the (weak) principle of equivalence as used by
Einstein. The  cosmological successes of General Relativity were
applications restricted to completely static systems for which
equation (\ref{LaueStatic}) holds.

John Norton has identified two other formulations of the principle
of equivalence, the first in Einstein's original writing and the
second in the work of those who based themselves on Einstein's
General Theory of Relativity. The first was expressed by Einstein
in 1918 and states that inertia and gravity are identical in
essence (wesensgleich) (\cite{Nortonb}, p. 233). This realist
version of the principle of equivalence seems to be an absolute
statement on the nature of gravity and inertia, independent of any
reference frame, and cannot be sustained in our present
interpretation. In the context of a Mie-de Broglie theory of
Quantum Gravity, inertia and gravity seem to be ""{\it
wesensungleich}"" (with double quotes, one for the language and
one to indicate the ambiguity of the use of "{\it wesen}" in our
case) because gravity seems to be a particle aspect of elementary
particles and inertia a wave aspect. Gravity and inertia exist
together as a fundamental and real particle-wave duality. This
duality is not an absolute statement but it is inferred or induced
from many experiments. The Harmony of the Phases makes this
duality compatible with the principle of relativity.

The second formulation was the infinitesimal formulation of the
principle of equivalence, attributed by Norton to Pauli and which
is now common in the context of the modern treatment of General
Relativity. It assumes the equality of inertial and gravitational
mass to hold only locally, in infinitesimal regions of space-time.
This version of the principle, also called the strong principle of
equivalence, was never accepted by Einstein (\cite{Nortonb}, p.
238). In Pauli's version the infinitely small world region
$\vartriangle \tau$ is so small that the space-time variation of
gravity is supposed to be negligible in it (\cite{Nortonb}, p.
235). The local version has been criticized not to be in
accordance with the appearance of tidal effects that do not vanish
inside the local cabin, however small it is made \cite{Nortonb}.
In our context, this version can be accepted in empty space where
all pressure vanishes and matter doesn't move, so applied to
completely static systems for which $\mathbf{v}\cdot\mathbf{g}=0$.
However, if this principle is used inside matter and in situations
with non-zero pressure, the infinitesimal principle can't be in
accordance with the basic empirical principles of Quantum
Mechanics. The strong principle just implies
$\mathbf{v}\cdot\mathbf{g}=0$ on a infinitesimal scale. This holds
in free space, but on a quantum scale in matter we have
Heisenbergs uncertainty relations
$\vartriangle\mathbf{p}\cdot\vartriangle\mathbf{r}\geq \hbar$ and
so $\mathbf{v}\cdot\mathbf{g}\vartriangle \tau \geq ic\hbar$ or
\begin{equation}
     \mathbf{v}\cdot\mathbf{g}= p_{quantum} \geq ic\frac{\hbar}{\vartriangle\tau}.
\end{equation}
On the infinitesimal scale of $\vartriangle\tau$ in matter, there
always is a quantum pressure because there always exists a non
zero action four-density. This implicates that wherever
Heisenbergs uncertainty relations practically limit the attainable
accuracy of measurements, the infinitesimal principle of
equivalence of inertial mass and gravitational mass becomes
invalid and only the equivalence of the phases may be used.

So in our interpretation, both the strong and the weak principle
of equivalence of the masses can be seen as the classical "limit"
of the principle of equivalence of the phases, when the
approximation $\hbar\approx0$ is valid {\it and} when inertial
wave-like clocks behave identical as gravitational inner-particle
clocks, that is when $\gamma\approx1$.

\section{Conclusion}

We believe that our interpretation based on the connection of
Mie's theory of matter and de Broglie's Harmony of the Phases
contains a new perspective on the old problem of the correct
formulation and use of the principle of equivalence. We do not
consider a scalar theory of gravity, as Mie's approach was, as
definitive. But if it is possible to integrate Quantum Physics to
a certain extend with a scalar theory of gravity, like Mie's, then
such a Scalar Quantum Gravity, however primitive, must contain
vital clues for the future development of a fully covariant
tensor-dynamical Quantum Gravity. The scalar theory of quantum
gravity can be developed further by connecting de Broglie's Theory
of the Double Solution, specially his attempt to connect the inner
clock-like frequency of particles to an inner warmth Q
(\cite{Sievers}, p. 40), to Mie's connection of thermodynamics and
gravity(\cite{Mie3}, p. 47-50), but for the time being we consider
that as a subject for future study. Last but not least we conclude
that if the connection made between Mie and de Broglie proves to
be physically valid, then a theory of Quantum Gravity based on the
orthodox interpretation of the Copenhagen School will be utterly
impossible. One cannot deny the existence of the particle in the
wave {\it and} connect something in that wave to the gravitational
energy! And the wave as defining only probabilities blocks the
view on its inertial properties needed in a theory of Quantum
Gravity. It seems to be an either/or situation, in which the
interpretation of the Copenhagen School dominates past and present
and Quantum Gravity has the future.

\section*{Acknowledgments:}
The author greatly appreciated the stimulating questions and
useful corrections of Dr. V. Dvoeglazov.

\end{document}